\documentclass[aps,prl,twocolumn,floatfix]{revtex4}

\usepackage{graphicx}
\usepackage{dcolumn}
\usepackage{bm}

\newcommand{\dfrac}[2]{\displaystyle\frac{#1}{#2}}

\newcommand{\kB}{k_{\text{B}}}
\newcommand{\rWS}{r_{\text{WS}}}
\newcommand{\thetaD}{\theta_{\text{D}}}
\newcommand{\thetaE}{\theta_{\text{E}}}

\begin{document}


\title{High pressure melt locus of iron from atom-in-jellium calculations}

\date{January 9, 2018, revisions to June 11, 2019
  -- LLNL-JRNL-769881}

\author{Damian~C.~Swift}
\email{dswift@llnl.gov}
\affiliation{%
   Lawrence Livermore National Laboratory,
   7000 East Avenue, Livermore, California 94551, USA
}
\author{Thomas~Lockard}
\affiliation{%
   Lawrence Livermore National Laboratory,
   7000 East Avenue, Livermore, California 94551, USA
}
\author{Raymond~F.~Smith}
\affiliation{%
   Lawrence Livermore National Laboratory,
   7000 East Avenue, Livermore, California 94551, USA
}
\author{Christine~J.~Wu}
\affiliation{%
   Lawrence Livermore National Laboratory,
   7000 East Avenue, Livermore, California 94551, USA
}
\author{Lorin~X.~Benedict}
\affiliation{%
   Lawrence Livermore National Laboratory,
   7000 East Avenue, Livermore, California 94551, USA
}

\begin{abstract}
Although usually considered as a technique for predicting electron states
in dense plasmas,
atom-in-jellium calculations can be used to predict the mean displacement of
the ion from its equilibrium position in colder matter, as a function of
compression and temperature.
The Lindemann criterion of a critical displacement for melting can then be
employed to predict the melt locus, normalizing for instance to the
observed melt temperature or to more direct simulations such as 
molecular dynamics (MD).
This approach reproduces the high pressure melting behavior of Al
as calculated using the Lindemann model and thermal vibrations in the solid.
Applied to Fe, we find that it reproduces the limited-range melt locus of
a multiphase equation of state (EOS) and the results of {\it ab initio} 
MD simulations,
and agrees less well with a Lindemann construction using an older EOS.
The resulting melt locus lies significantly above the older melt locus
for pressures above 1.5\,TPa, but is closer to recent {\it ab initio} MD 
results and extrapolations of an analytic fit to them.
This study confirms the importance of core freezing in massive exoplanets,
predicting that a slightly smaller range of exoplanets than previously assessed
would be likely to exhibit dynamo generation of magnetic fields by convection
in the liquid portion of the core.
\end{abstract}


\maketitle

\section{Introduction}
Thousands of exoplanets have been discovered \cite{exoplanets},
most around stars of different types than the sun, and with orbits and
mean mass density of a much wider variety than the planets of the solar system.
These observations lead to questions about the uniqueness of the solar system
and the Earth, including whether other planets can support life.

All known forms of life require liquid water (though extremophiles can survive
frozen or even in vacuum when dormant \cite{Rothschild2001}),
and almost none can tolerate ionizing radiation at the levels typical
of the solar wind and flares.
Earth's magnetic field shields the atmosphere from energetic charged particles,
and so a magnetic field is usually considered a prerequisite for life 
\cite{sasselov12}.

Earth's magnetic field is believed to be induced by convection in the 
liquid Fe outer core \cite{dehant07}, 
therefore an important indication of the habitability
of rocky `super-Earth' exoplanets is whether the core is likely to possess
a liquid layer.
This depends on the circumstances of each particular exoplanet, including
its composition -- influencing the specific Fe alloys in the core
as well as the proportion of silicates to Fe -- and history, which
depends on the type of star it orbits and also interactions with other
exoplanets in the system,
but the relevent material physics property is the melt locus of Fe.

A large number of exoplanets have been observed with mass and radius 
indicating rocky structures analogous to Earth \cite{sasselov08},
and there is an increasing body of research predicting whether they are
likely to contain liquid Fe in the core, assuming compositions similar to
Earth \cite{valencia06,sotin07,gaidos10,Stixrude2014}.
Even neglecting the variation and uncertainty in composition,
different studies have reached inconsistent conclusions because of our 
uncertain knowledge of the properties of Fe at elevated pressures and
temperatures, in particular the relationship between the planetary
temperature profile and the melt loci of the core and mantle 
\cite{sotin07,valencia09}.
The temperature profile in the Earth's core crosses 
the melt locus of Fe at $\sim$330 GPa \cite{dziewonski71,dziewonski81}. 
The magneto-dynamo is thought to be driven by the latent heat of solidification
as the inner core grows \cite{gubbins11}, and may be affected by the
expulsion of lighter impurity elements such as Si and S from the solid;
the impurity composition is thought to vary even between the rocky planets
of our solar system \cite{mercurycomp,marscomp}.
Conclusions vary between inferring that planets larger than Earth would have 
a completely solid core and hence no magnetic field \cite{valencia06}
to predictions that a liquid outer core could be present in planets
up to five times the mass of the Earth \cite{sotin07}.
Theoretical predictions of the melt locus of Fe \cite{Morard2011,bouchet13}
lie significantly higher than Lindemann law extrapolations from low pressure
data \cite{valencia06}, suggesting a smaller possible population of
super-Earths with a magnetic field.
However, these conclusions depend on the temperature profile, which depends also
on the properties of the mantle \cite{Stixrude2014}.
As well as indicating a wider range of occurrence of liquid Fe, it has been
suggested that convection in the core could be driven alternatively by
convection in the mantle \cite{Stixrude2014}.

The melt locus is defined most rigorously by thermodynamic construction,
matching the Gibbs free energy of the liquid and solid phases.
The EOS of solid phases can be determined theoretically using electronic
structure to infer the free energy as a function of mass density and
temperature, which may be decomposed as a cold compression curve
plus phonon modes and possibly electron excitations.
These contributions can be calculated from static lattice simulations,
although in some cases it has been shown that the phonon and electron
excitations may interact significantly, necessitating temperature-dependent
corrections such as anharmonic phonons \cite{Stixrude2012}.
The equivalent calculations for the free energy of the fluid require
quantum molecular dynamics (QMD), in which the 
kinetic motion of an ensemble of atoms is simulated, with the instantaneous
forces on the ions found from electronic structure calculations \cite{qmd}.
QMD can also be used to calculate the EOS of the solid directly,
which automatically incorporates interactions between electronic and ionic
excitations, 
but, since the ion motion is classical, makes it more difficult 
to account for zero-point motion of the ions.
QMD can be used to deduce the melt locus directly, by performing simulations
comprising regions of solid and fluid in contact.
In such simulations, one phase grows at the expense of the other,
and each point on the melt locus is identified by adjusting the state until
the interface is approximately stationary.
The state in the simulation can be adjusted until the phases remain in
equilibrium, identifying a point on the melt locus.
Either procedure is computationally expensive, requiring $o(10^{16})$ or more
floating-point operations to identify a state on the melt locus, 
equivalent to thousands of CPU-hours per state.
These calculations are typically more expensive at lower compressions,
and it is common for studies to focus only on a narrow range of pressures.

As melt loci are challenging to predict theoretically,
particularly over a wide pressure range,
many studies rely on melt loci deduced much more simply,
such as by the Lindemann criterion applied to wide-range semi-empirical
EOS \cite{Steinberg1996,sesrefs,qeos,leos}.
Typically, an EOS is constructed using adjustable models of the cold curve
and ion-thermal excitations, and the melt locus is constructed from the
Lindemann law with the displacment criterion chosen to pass through 
available melting data, such as the observed melt temperature at 1\,atm.
In practice,
the Lindemann law is solved as a first-order differential equation in 
mass density, relating the melt temperature to the ion-thermal
Gr\"uneisen parameter \cite{Lindemann1910,Gilvarry1956}.
Usually, the highest-compression data available lie along the principal
shock Hugoniot. The split between cold and thermal pressure is not
constrained, and the melt locus at high pressures therefore depends 
on the extrapolation of empirical functions for cold curve, Debye temperature,
and Gr\"uneisen parameter.
It is also possible to estimate the ion-thermal free energy from the
cold curve, for instance by estimating the Debye temperature from the
bulk modulus \cite{Moruzzi1988}, although this approach has been found to
become less accurate as pressure increases;
the melt curve can then be estimated again using the Lindemann criterion
\cite{Steinberg1996}.
Such EOS and melt loci are used in the design and interpretation 
of expensive high energy density experiments
such as those at the National Ignition Facility, 
which often rely on predictions of whether or not components melt
\cite{nifrefs}.

Recent QMD and path-integral Monte Carlo 
results have indicated that the simpler approach of
calculating the electron states for a single atom in a spherical cavity
within a uniform charge density of ions and electrons, representing the
surrounding atoms, reproduces their more rigorous EOS for dense plasmas
\cite{Benedict2014,Driver2017}.
This atom-in-jellium approach \cite{Liberman1979} was developed originally
to predict the electron-thermal energy of matter at high temperatures and
compressions \cite{sesrefs}, as an advance over the primitive electronic
models neglecting any treatment of shell structure,
as in Thomas-Fermi and related approaches \cite{tf}.
A development of atom-in-jellium was used to predict ion-thermal properties
\cite{Liberman1990},
and seems to give reasonable EOS in the fluid regime down at least as far as
the melt locus for a wide range of elements \cite{Swift2018_aahedeos}.
In a further development of the method, we have used Hellmann-Feynman 
calculations of the restoring force for perturbations of the ion from its
equilibrium position to predict the transition from bound to free ions,
resulting in a reduction in the ion-thermal heat capacity from $3 \kB$ per atom
to $3 \kB/2$ per atom, by considering the mean displacement of the ions
\cite{Swift2019_aaion}.

In the work reported here, we use the atom-in-jellium displacement model 
developed previously to estimate the melt locus of elements efficiently
over a wide range of pressures, and assess the astrophysical implications 
compared with previous melt locus calculations for Fe.

\section{Atom-in-jellium ionic displacement model}
In the ion thermal model developed for use with atom-in-jellium calculations
\cite{Liberman1990},
perturbation theory was used to calculate the Hellmann-Feynman force on the
ion when displaced from the center of the cavity in the jellium.
Given the force constant $k=-\partial f/\partial r$, the Einstein vibration
frequency $\nu_e=\sqrt{k/m_a}$ was determined, where $m_a$ is the atomic mass,
and hence the Einstein temperature $\thetaE=\hbar\nu_e/\kB$.
The Debye temperature $\thetaD$ was inferred from $\thetaE$, either by
equating the ion-thermal energy or the mean displacement $u$.
We used the mean fractional displacement with respect to the 
Wigner-Seitz radius, $u_f\equiv u/\rWS$, 
as a measure of ionic freedom,
describing the decrease in ionic heat capacity from 3 to $\frac 32 \kB$
per atom as the ions become free as temperature increases in the fluid
\cite{Swift2019_aaion}.

Predictions of the variation of $\thetaD$ with temperature as well
as density are unusual compared with the normal use in constructing EOS.
This behavior adds generality, and is likely to make a Debye-based EOS
construction valid over a wider range of states.
The treatment of ionic freedom extended the atom-in-jellium technique
to describe the variation of the ion-thermal heat capacity into regimes
where the electronic treatment is more appropriate.

The use of fractional displacements to predict ionic freedom
is reminiscent of the semi-empirical Lindemann melting criterion 
\cite{Lindemann1910}, which holds that melting occurs when the mean displacement
of the atoms reaches some fixed fraction of the interatomic spacing.
This fraction is approximately constant for a given material but varies
somewhat with composition; 
the value is found to vary between around 0.1 and 0.3.
The variation in fractional density change from solid to liquid
is much less than this range,
so it is reasonable to perform the same calculation using the
interatomic spacing in the liquid rather than the solid.

With such a wide variation in fractional displacements inferred to
induce melting, this method is not predictive by itself.
For substances with a simple phase diagram, melting at one atmosphere
could be used to constrain the critical value of $u_f$.
For substances that exhibit multiple solid phases with significant 
volume changes, the melt locus is typically perturbed by the free energy 
variations in the solid.
Thus we would anticipate that the atom-in-jellium displacement technique
could capture the variation in melt locus for melting from each solid
phase away from the phase boundaries, but would require normalization
for each solid phase, e.g. to QMD simulations.

However, the atom-in-jellium calculations are much faster than QMD,
so a relatively small number of QMD simulations could be used to constrain
finely-resolved and wide-ranging atom-in-jellium calculations.
Furthermore, the atom-in-jellium calculations
are fast enough that all electrons can be treated explicitly
under all circumstances, 
in contrast to QMD simulations where the inner electrons are
typically subsumed into a pseudopotential;
atom-in-jellium calculations can be used to extrapolate to much lower and higher
densities than are tractable with QMD.

\section{Melt locus of aluminum}
\begin{figure}
\begin{center}\includegraphics[scale=0.75]{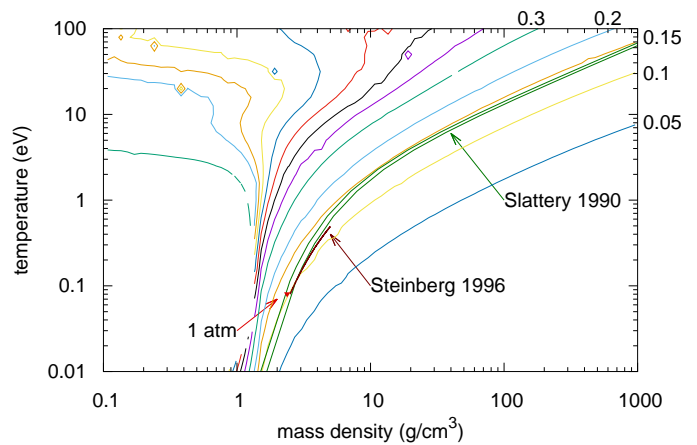}\end{center}
\caption{Contours of mean fractional displacement, and melt loci for aluminum
   extracted from EOS models \cite{Steinberg1996,Slattery1990}.
   Melt loci and one atmosphere melting are indicated.
   The other curves are contours of mean fractional displacement,
   with relevant values marked.}
\label{fig:Almelt}
\end{figure}

At atmospheric pressure,
Al melts at 933.47\,K with a liquid density 2.375\,g/cm$^3$
\cite{CRC2011,CRC2003}.
Modified Lindemann melt loci have been developed to be consistent with
an analytic Gr\"uneisen EOS \cite{Steinberg1996} and a wider-ranging
tabular EOS \cite{Slattery1990};
the melt loci were consistent with each other.
Atom-in-jellium calculations were performed for
$10^{-4}$ to $10^3\rho_0$ with 20 points per decade,
and $10^{-3}$ to $10^5$\,eV with 10 points per decade.
A non-imaginary Einstein frequency was calculated for $\rho>1.3$\,g/cm$^3$, 
indicating that the atoms were not free,
allowing the melt locus to be estimated in this range.
The melt loci corresponded to $u_f\simeq 0.1$ for one atmosphere melting,
rising to $\simeq 0.13$ at 5\,g/cm$^3$ (1\,eV), 
and then following within 0.01 of this displacement contour
to the maximum density considered.
(Fig.~\ref{fig:Almelt}.)

This result suggests that fractional displacements $u_f$ 
derived from atom-in-jellium
calculations can be used to predict the melt locus, with deviation
from a constant $u_f$ becoming significant around ambient density where the
inaccuracies in atom-in-jellium electronic states result also in significant
inaccuracies in the EOS.

\section{Melt locus of iron}
\begin{figure}
\begin{center}\includegraphics[scale=0.75]{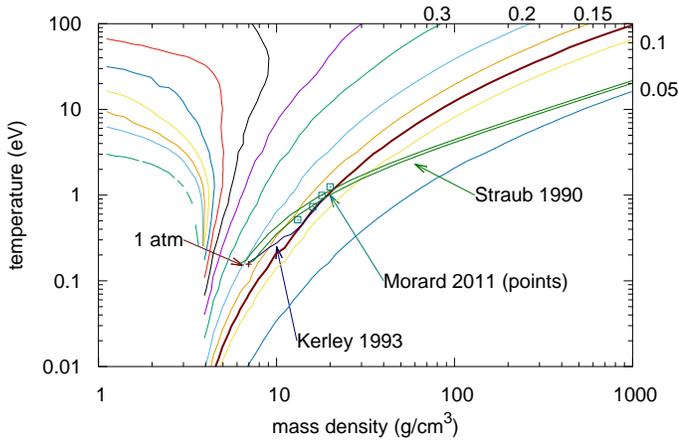}\end{center}
\caption{Contours of mean fractional displacement, melt loci for iron
   extracted from EOS models \cite{Straub1990,Kerley1993},
   and QMD melting predictions \cite{Morard2011}.
   Bold contour is atom-in-jellium locus at mean fractional displacement of 0.12,
   proposed as improved high pressure melt locus.}
\label{fig:Femelt}
\end{figure}

Because of the solid-solid phase transitions in Fe,
we would not expect the
atom-in-jellium fractional displacements to be constant between and around
the triple points with the liquid.
A wide-ranging Lindemann melt locus was developed to be consistent with
a tabular EOS that did not treat the solid phase transitions
\cite{Straub1990}.
A multiphase EOS was developed describing and extrapolating from
experimental measurements, and treating solid phases
$\alpha$, $\gamma$, $\epsilon$, and $\delta$ as well as the liquid/vapor
region \cite{Kerley1993}.
Phase boundaries were extracted from this tabular EOS by locating anomalies
in heat capacity and Gr\"uneisen parameter; the melt transition was evident
up to a density around 20\,g/cm$^3$.
QMD studies of the melt locus have also been performed at 13-20\,g/cm$^3$
\cite{Morard2011}.
The multiphase melt locus was similar to the QMD results;
the older melt locus passed through the others around 20\,g/cm$^3$
but varied significantly more slowly with compression.
Atom-in-jellium calculations were performed over the same compression and
temperature range as for Al, and predicted a physical Einstein temperature for 
$\rho>4$\,g/cm$^3$.
One atmosphere melting corresponds to $u_f\simeq 0.17$.
Above 12\,g/cm$^3$,
the multiphase and QMD simulations were very close to the $u_f=0.12$ contour.
Thus we propose the $u_f=0.12$ contour as an improved melt locus to higher
pressure.
(Fig.~\ref{fig:Femelt}.)

Multi-atom electronic structure calculations have been used to
predict solid phase stability in Fe at high pressure \cite{Stixrude2012}.
These indicated a transition from hcp to fcc at 33.9\,g/cm$^3$ (6\,TPa),
and fcc to bcc at 66.7\,g/cm$^3$ (38.3\,TPa).
The energy difference between hcp and fcc is relatively small,
so this transition is unlikely to affect the melt locus much.
The transition to bcc could be associated with an increase in the slope of
the melt locus.

\begin{figure}
\begin{center}\includegraphics[scale=0.72]{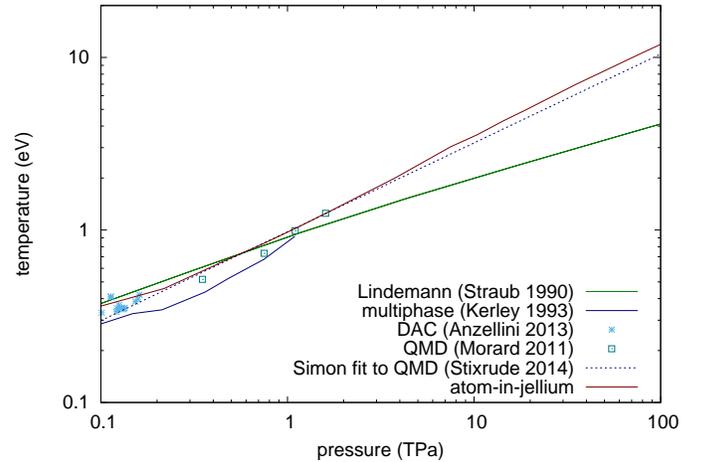}\end{center}
\caption{Melt loci for iron as a function of pressure,
   also showing heated diamond anvil cell (DAC) measurements at low pressures
   \cite{Anzellini2013}.}
\label{fig:Femeltp}
\end{figure}

Taking the set of $\{\rho,T\}$ points along the contour, thermodynamic
quantities were calculated for the melt locus by interpolation from the
atom-in-jellium EOS.
Again, because of the relative inaccuracy of the atom-in-jellium method at
densities near ambient, one would not expect the pressure to be accurate
in this regime.
However, the pressure was found to match the QMD simulations to within 10\%\,
and is likely to be at least as accurate at higher pressures.
The atom-in-jellium melt locus is significantly higher than 
a Lindemann-based prediction
using plane-wave pseudopotential calculations of vibrational frequencies
\cite{Stixrude2012},
but is reasonably consistent with the same researcher's
Simon fit to the QMD locus \cite{Stixrude2014}.
The latter extrapolates from the QMD calculations, which define
four points ranging $\sim$0.35-1.5\,TPa with uncertainties 
in temperature which, strictly,
give a significant variation in the extrapolation to higher pressures.
The atom-in-jellium calculation provides some validation of this 
extrapolated melt locus, but rises above it by 10\%\ at a pressure of 10\,TPa,
suggesting that a slightly smaller proportion
of exoplanets are likely to possess a magnetic field induced
by convection in the core.
(Fig.~\ref{fig:Femeltp}.)

The atom-in-jellium melt locus could not be fit over its full range
using a function of the Lindemann or Simon type.
Over the range 1-100\,TPa thought to be most relevant to giant exoplanets
\cite{Stixrude2012}, the melt locus could be reproduced 
to within 3\%\ by the Simon equation \cite{Simon1927}
with parameters as follows:
\begin{equation}
T_m=6279\,\text{K}\left(\dfrac p{346\,\text{GPa}}\right)^{0.552}.
\end{equation}
The range of the fit can be broadened to 0.5-650\,TPa by modifying the
exponent:
\begin{equation}
T_m=494\,\text{K}\left(\dfrac p{3\,\text{GPa}}\right)^{0.543-p/(4\times 10^7\,\text{GPa})}.
\end{equation}

\section{Discussion}
The atom-in-jellium model was developed for application to warm dense matter,
and it is surprising that it can be used to predict melt loci.
To emphasize, this is a generalization of the Lindemann model,
based on the empirical observation that melting occurs at a roughly
constant mean displacement of the ions from equilibrium, rather than any
more rigorous representation of the free energy difference between the
solid and liquid.
As used here, predicted melt loci are also subject to the inherent 
approximations of the average-atom treatment and of the first order perturbation
approach to calculating the Einstein temperature.
However, the procedure used here for calculating the melt locus is
significantly different than previous approaches:
rather than integrating an equation involving the ion-thermal 
Gr\"uneisen parameter
(which was not even calculated in constructing the
atom-in-jellium EOS, although it can be deduced from the
ionic component of the EOS by differentiation),
the mean amplitude of vibrations used in computing the Debye frequency 
was used directly to determine the melt locus,
with no integration required
and thus no accumulation of error with increasing compression.
The atom-in-jellium melt loci agree encouragingly well with
experimental measurements and more rigorous calculations over the narrower
ranges where they exist,
and are likely to be more accurate than extrapolations of 
empirical EOS or melt constructions,
and so should be useful for high pressure situations including 
massive exoplanets, white and brown dwarfs, and high energy density experiments.

The melt locus proposed here confirms the previous conclusion
\cite{Stixrude2014} that the size of the frozen core of
Fe planets should grow monotonically with planetary mass, at least 
for planets of broadly constant composition.
The frozen core would grow much less, and possibly shrink,
depending on the scenario assumed, using the
earlier melt locus for Fe \cite{Straub1990}.
This observation highlights the potential importance of convection in
the mantle as a mechanism for generating magnetic fields in massive
or silicate-rich exoplanets.
This modified melt locus is important in assessing whether specific,
detailed scenarios of planetary formation and evolution are
potentially compatible with the occurrence of extraterrestrial life.

\section{Conclusions}
Einstein oscillator estimates from the atom-in-jellium model of
warm dense matter were used to calculate the mean thermal displacement
of ions as a function of mass density and temperature.
Expressed as a fraction of the Wigner-Seitz radius as a measure of
interatomic spacing,
contours of this fractional displacement were found to reproduce
experimental measurements and more rigorous calculations of the melt locus
of Al and Fe, except near solid-solid phase transitions.
Having established the mean fractional displacement corresponding to
melting, the calculated contour can be used to predict the melt locus
to much higher pressures with a sounder physical basis than
extrapolations based on empirical fits to the EOS of the solid and liquid
phases or to the melt locus itself,
which are the methods generally used.

For Fe, the atom-in-jellium melt locus broadly confirms and refines
a recent prediction based on an empirical extrapolation of QMD calculations.
This result shows
the importance of the high pressure melt locus to the range of
conditions in which convection can occur in the core of massive exoplanets,
and therefore in which magnetic fields can be generated by the core dynamo 
process, with implications for the population of candidate life-bearing planets.

\section*{Acknowledgments}
The authors would like to thank Richard Kraus for useful discussions.
J.D.~Johnson and Scott Crockett provided 
copies of the {\sc sesame} equation of state library.
G.I.~Kerley provided a copy of his equation of state for Fe.

This work was performed under the auspices of
the U.S. Department of Energy under contract DE-AC52-07NA27344.

\end{document}